\documentclass[twocolumn,english,superscriptaddress,floatfix,longbibliography,nofootinbib]{revtex4-2}

\usepackage{graphicx}
\usepackage{dcolumn}
\usepackage{bm}

\usepackage[utf8]{inputenc}
\usepackage[T1]{fontenc}
\usepackage{mathptmx}
\usepackage{etoolbox}

\usepackage{braket}
\usepackage{subcaption}
\usepackage{ragged2e}
\DeclareCaptionJustification{justified}{\justifying}
\usepackage{booktabs}
\usepackage{amsmath}

\newcommand{\curlE}{\mathcal{E}}
\newcommand{\ii}{\text{i}}

\usepackage{placeins}
\usepackage{xcolor}
\definecolor{mygreen}{rgb}{0.328125,0.6796875,0.1953125}
\definecolor{myblue}{rgb}{0.12156862745098039, 0.4666666666666667, 0.7058823529411765}
\usepackage{hyperref}
\hypersetup{
     colorlinks=true,
     linkcolor=myblue,
     filecolor=magenta,      
     urlcolor=myblue,
     pdftitle={Magic state distillation without measurements and post-selection},
     citecolor=myblue
}

\makeatother
\begin{document}

\title{Magic state distillation without measurements and post-selection}
\author{Sascha Heußen}
\email{s.heussen@neqxt.org}
\affiliation{neQxt, 50670 Cologne, Germany} 

\date{\today}

\begin{abstract}
Magic state distillation (MSD) is a quantum algorithm that enables performing logical non-Clifford gates with in principle arbitrarily low noise level. It is herein typically assumed that logical Clifford gates can be executed without noise. Therefore, MSD is a standard subroutine to obtain a fault-tolerant universal set of quantum gate operations on error-corrected logical qubits. Well-known schemes conventionally rely on performing operator measurements and post-selection on the measurement result, which makes distillation protocols non-deterministic in the presence of noise. In this work, we adapt the 15-to-1 MSD protocol such that it deterministically suppresses noise by using a coherent feedback network on the output states without the need to perform individual qubit measurements. These advantages over textbook MSD come at the price of reducing the noise suppression per round from $\mathcal{O}(p^3)$ to $\mathcal{O}(p^2)$. Our technique can be applied to any MSD protocol with an acceptance rate of 1 in the absence of noise. It may be desirable to use our scheme if the coherent feedback network can be executed faster and more reliably than the measurements and/or if logical clock cycles in the quantum processor should be kept synchronous at all times. Our result broadens the path of potential experimental realizations of MSD in near-term devices and advances the development of fault-tolerant quantum computers with practical use.
\end{abstract}

\maketitle

\textbf{Introduction.}
In order to unleash the potential of quantum computers, qubits should be equipped with a universal set of quantum gate operations \cite{kitaev1997quantum, solovay2000lie}. While this -- in theory -- enables the computationally efficient approximation of any unitary operation with arbitrary precision, today's quantum computing hardware is yet too noisy to reliably execute the large number of gates believed to be necessary for an algorithmic quantum advantage with practical use. Quantum error correction (QEC) is a framework for the realization of fault-tolerant quantum computation that holds the promise to execute large-scale algorithms by suppressing noise to in principle arbitrarily low levels \cite{devitt2013quantum}. Fault-tolerant (FT) gate operations need to be performed directly on the encoded, error-corrected, qubits in order to uphold the code's protection during a computation \cite{preskill1998reliable, campbell2017roads}. It is desirable to use QEC codes with a large set of transversal gates because transversal gates are inherently FT without additional qubit or gate overhead. Yet, no QEC code can have a universal transversal gate set and therefore alternative FT gate constructions are required \cite{eastin2009restrictions}.

Magic state distillation (MSD) algorithms sequentially improve the fidelity of certain resource quantum states that are used, after the distillation, to enact a logical quantum gate operation on an arbitrary qubit state \cite{bravyi2005universal}.
Assuming that Clifford gates are easily implemented fault-tolerantly, for example via a QEC code with transversal Clifford gates, a distilled magic state can be input to a gate teleportation circuit: This "noiseless" circuit, as a result, performs a high-fidelity logical gate, for example a non-Clifford gate for said QEC code, on an arbitrary input state and the magic state is consumed in the process. 
In this framework, the noise-level of the magic state determines the noise-level of the logical gate and can be controlled systematically via the number of MSD rounds. 

Previously, MSD has been thought to pose an unfeasibly large qubit and gate overhead for current quantum processors. Magic state injection, code switching and magic state cultivation have evolved as alternatives for FT universal gate set realizations \cite{goto2016minimizing, chamberland2020very, butt2024fault, heussen2024efficientfaulttolerantcodeswitching, gidney2024magicstatecultivationgrowing}. A crucial aspect that limits the practical applicability of MSD in real quantum computing hardware is the consistent difficulty to perform fast and accurate measurements with real-time feed-forward as part of a quantum circuit. Therefore, there has been a growing interest in measurement-free FT circuits for QEC and logical gate operations \cite{paz2010fault, heussen2024measurement, veroni2024optimized, butt2024measurementfreescalablefaulttolerantuniversal, veroni2025universalquantumcomputationscalable}. Recently, however, we have witnessed first small-scale experimental realizations of MSD in ion traps and neutral atom platforms that use the relatively small \mbox{5-to-1} MSD protocol on logical qubits that are encoded in two-dimensional color codes \cite{brown2023advancescompilationquantumhardware, rodriguez2024experimentaldemonstrationlogicalmagic}. 

Since its introduction, MSD has evolved into its own field of research dealing with distillation of different types of magic states that facilitate the teleportation of various logical gates, considering combinations of different schemes into especially effective distillation sequences, exploring the mathematical details of distillable states or scrutinizing the efficiency of implementation in near-term hardware \cite{reichardt2004improved, campbell2009structure, campbell2010bound, meier2012magic, jochymoconnor2012robustnessmagicstatedistillation, bravyi2012magic, jones2013multilevel, eastin2013distilling, campbell2017unified, haah2017magic, o2017quantum, howard2017application, haah2018codes, hastings2018distillation, litinski2019magic, gidney2019efficient, haah2021measurement, bao2022magic, lee2025lowoverheadmagicstatedistillation, kalra2025invarianttheorymagicstate, fazio2025lowoverheadmagicstatecircuits}. The authors of Ref.~\cite{beverland2021cost} aim to quantify the overhead required to reach fault-tolerant universality and provide an accessible high-level overview of known magic state distillation schemes. 

In this work, we contribute a new circuit implementation for the 15-to-1 magic state distillation scheme, which falls into class A of the terminology of Ref.~\cite{beverland2021cost}. The 15-to-1 scheme is based on the $[[15,1,3]]$ QEC code and, for being class A, yields a suppression of noise from $\mathcal{O}(p)$ to $\mathcal{O}(p^3)$ with an acceptance rate that approaches unity in the low-noise limit. We report that, for such protocols, one need not reject the result of MSD when the scheme indicates so, but may instead use a look-up-table decoder of the corresponding QEC code to flip the output magic state depending on the syndrome. Doing so reduces the noise suppression to $\mathcal{O}(p^2)$. Another desirable effect, however, is that these correction operations can be performed without measurements but coherently, with the help of additional multi-qubit-controlled gates. As a result, our scheme enables the experimental realization of magic state distillation without the need to post-select on measurement results and without performing measurements at all. This renders measurement-free MSD feasible for a broader range of existing quantum processors \cite{kaufman2021quantum, graham2022multiqubit, scholl2023erasure, ma2023highfidelity, bluvstein2024logical, lekitsch2017blueprint, kaushal2020shuttling, pogorelov2021compact, monroe2013scaling, moses2023race, yu2025insitumidcircuitqubitmeasurement, chen2025noninvasivemidcircuitmeasurementreset} without compromising on the original idea to sequentially distill magic states until a desired accuracy is reached. Our scheme may favorably integrate into FT quantum computing architectures because the distillation time is a priori determined: It allows one to produce magic states in a synchronized fashion with logical clock cycles as there is no need to non-deterministically wait until a certain round of MSD has finished. 

This work is structured as follows. We subsequently introduce unitary encoding and decoding circuits that are used to obtain syndromes of noisy magic states with a distillation QEC code. Thereupon, we show how a coherent feedback network can be constructed systematically for the 15-to-1 MSD protocol. Moreover, we provide numerical simulation results that demonstrate the effectiveness of our measurement-free MSD implementation and indicate that the analytically expected scaling behavior is achieved up to relatively high error rates. Lastly, we draw conclusions and give an outlook toward future work. 

\textbf{Unitary decoding circuits.}
Magic state distillation protocols are conventionally performed by measuring stabilizer operators of a suitable QEC code \cite{devitt2013quantum}. A large number of more-noisy magic states are transformed via these multi-qubit operator measurements, which are prescribed by the MSD scheme, into a smaller number of less-noisy magic states via non-deterministic post-selection on a certain syndrome measurement outcome.

This is not the only way to determine the syndrome of a noisy logical qubit state. In fact, since we do not aim to denoise a logical qubit state in order to extend its lifetime (as for QEC cycles), we can employ unitary encoding and decoding circuits of an $[[n, k, d]]$ QEC code to map an initial quantum state $\ket{\psi_1,\psi_2,...,\psi_k}$ on $k$ physical \emph{message} qubits to an $n$-qubit logical state $\ket{\overline{{\psi_1,\psi_2,...,\psi_k}}}$ and vice versa using only unitary gate operations. Such circuits can be constructed systematically, for instance, with the help of ZX-calculus \cite{kissinger2022phasefreezxdiagramscss}.

We here consider a particular class of distillation procedures, where a logical magic state is prepared by performing a logical gate operation on an encoded qubit via $n$ noisy input magic states. Then, the \emph{decoding} circuit distills these states by mapping the logical magic state(s) on $n$ physical qubits back to $k$ physical qubits. Therefore, these schemes always succeed in the absence of noise. The remaining $n-k$ physical output qubits carry the syndrome information. While they might be used to post-select on the trivial syndrome, i.e.~the state $\ket{0}^{\otimes n-k}$, the available information may be sufficient to infer a correction operation on the $k$ message qubits. 

\begin{figure}\centering
	\includegraphics[width=0.99\linewidth]{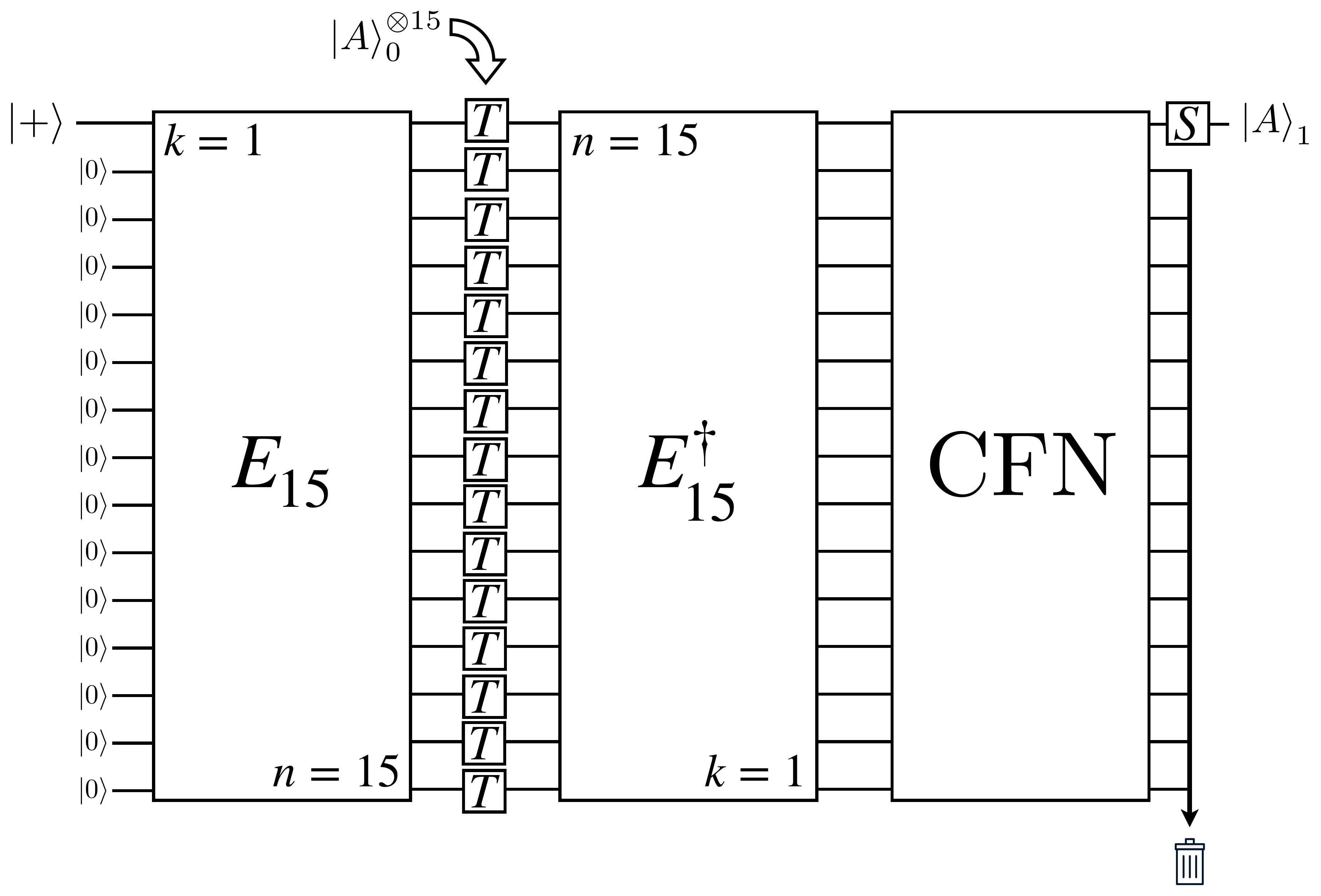}
	\caption{Magic state distillation (MSD) of the state \mbox{$\ket{A} = \frac{1}{\sqrt{2}}\left(\ket{0} + e^{\ii \pi/4}\!\ket{1}\right)$} using the unitary encoding map $E_{15}$ of the $[[15,1,3]]$ code in reverse. The logical $T$-gate is transversal in the $[[15,1,3]]$ code. A physical $T$-gate can be implemented via a gate teleportation circuit that consists of only Clifford gates and one input magic state $\ket{A}_0$. The $n=15$ input magic states are mapped onto $k = 1$ output magic state $\ket{A}_1$. We subsequently apply a coherent feedback network (CFN) and then discard the other 14 output qubits. The circuit boxes are laid out explicitly in Fig.~\ref{fig:physical_circ}.}
	\label{fig:mdist151}
\end{figure}

Our MSD scheme is pictured in Fig.~\ref{fig:mdist151}. Here, we focus on the 15-to-1 MSD protocol, which makes use of the $[[15,1,3]]$ QEC code \cite{steane1999quantum, bombin2015gauge} to distill $n = 15$ to $k=1$ magic states. This code can correct a single arbitrary Pauli error and it can detect up to two arbitrary Pauli errors because it has distance $d=3$. It furthermore has a transversal logical $T$-gate and is capable of distilling the magic state \mbox{$\ket{A} = \frac{1}{\sqrt{2}}\left(\ket{0} + e^{\ii \pi/4}\!\ket{1}\right)$}. This state facilitates a $Z$-rotation about an angle of $\pi/4$ when used in gate teleportation.

\textbf{Coherent feedback network.}
For MSD schemes that utilize distance-$d$ QEC codes to suppress noise of strength $\mathcal{O}(p)$ to $\mathcal{O}(p^d)$ in post-selection, the feedback prescribed by the QEC decoder may be used to deterministically perform correction operations based on the syndrome information. Thereby, the noise suppression is reduced to $\mathcal{O}(p^{\lfloor \frac{d+1}{2} \rfloor})$ since this is the leading order in $p$ where uncorrectable errors occur in the low-$p$ limit. For small QEC code instances, most-likely-error (MLE) decoding can be performed in practice despite its exponential scaling properties. In-sequence stabilizer measurements and feed-forward corrections based on classical, pre-computed look-up tables can be replaced with a fully coherent feedback network (CFN) based on quantum gate operations \cite{heussen2024measurement, veroni2024optimized}. We stress that, in the context of distillation, not \emph{all} correctable errors\footnote{There are $2^{n-k}-1$ different non-trivial syndromes for an $[[n,k,d]]$ QEC code. For this reason, MLE decoders such as look-up tables grow exponentially with the size of the used code. The CFN may therefore add an exponential number of gates and complicate scale-up when used for QEC cycles where \emph{all} $s \leq t = \lfloor d-1/2 \rfloor$ errors need to be corrected.} on the $n$-qubit state actually need to be detected and distinguished but \emph{only} the errors that end up on the message qubits and, as a consequence, lead to an erroneous output state.

To incorporate measurement-free feedback operations into our 15-to-1 MSD protocol, we design a CFN that only corrects those errors, which proliferate through the unitary decoding circuit and harm the output qubit. As a first step, we use Pauli propagation rules \cite{gottesman1998heisenbergrepresentationquantumcomputers} for every single-qubit Pauli operator on an input wire of the decoding circuit to 1) determine whether the message qubit is flipped or remains unaffected and 2) record the corresponding state of the syndrome qubits. The exhaustive list of possible syndromes is given alongside the physical quantum circuit in App.~\ref{sec:circ}. We then use this list, shown in Tab.~\ref{tab:zx-sector-flips}, to design a sequence of multi-qubit-controlled gates where the controls are wired to (a subset of) the syndrome qubits and the target is wired to the output message qubit, as depicted in Fig.~\ref{fig:physical_circ}. The gate combination is carefully crafted such that the message qubit is flipped when a corresponding syndrome is the result of the unitary decoding circuit and left unchanged otherwise. Note that, in this setting, there are no Clifford corrections associated with commuting a Pauli error past a multi-qubit-controlled gate (as for example in Ref.~\cite{chao2018fault}). This is due to the fact that the $n-k$ syndrome qubits that control the correction operations can only be in computational basis states $\ket{0}$ or $\ket{1}$ after running $E_{15}^\dag$. 

Such a CFN seems convenient for application on logical qubits that are encoded in a QEC code with a suitable FT construction of multi-qubit-controlled gates; see for instance the construction of FT $C^kZ$ gates in the Bacon-Shor code given in Refs.~\cite{yoder2017universalfaulttolerantquantumcomputation, veroni2025universalquantumcomputationscalable}. Although these gates are not simply transversal, the authors suggest FT constructions by consecutively extending and shrinking code patches for arbitrary code distances. In Fig.~\ref{fig:physical_circ_cr}, we additionally provide a simplified implementation based on coherently-controlled-reset operations that does not use multi-qubit-controlled gates.

\textbf{Measurement-free distillation circuit analysis.}
Let us now scrutinize the performance of our 15-to-1 MSD scheme in the presence of noise.
For our numerical simulations we use a modified version of \texttt{PECOS} \cite{pecos}. 

It is conventional to assume a twirled noise model where input magic states are stochastically flipped to their orthogonal partner state, i.e., the state $\ket{A}$ is flipped with a probability $p$ to the orthogonal state $\ket{A_\perp} = Z \ket{A}$ \cite{bravyi2005universal}. Any noisy magic state in the MSD protocol is then described by a single-qubit density matrix
\begin{align}
    \rho = \curlE(\ket{A}\bra{A}) &= (1-p)\ket{A}\bra{A} + p \ket{A_\perp}\bra{A_\perp} \notag \\
    &= (1-p)\ket{A}\bra{A} + p\,Z \ket{A}\bra{A} Z \label{eq:noise}.
\end{align}
Gates are assumed to be noise-free as the result of employing MSD on logical qubits that are encoded in an $[[N, K, D]]$ QEC code\footnote{Note that this logical qubit QEC code is, in general, different from the distillation QEC code.}
in the limit $D \rightarrow \infty$.

\begin{figure}\centering
	\includegraphics[width=\linewidth]{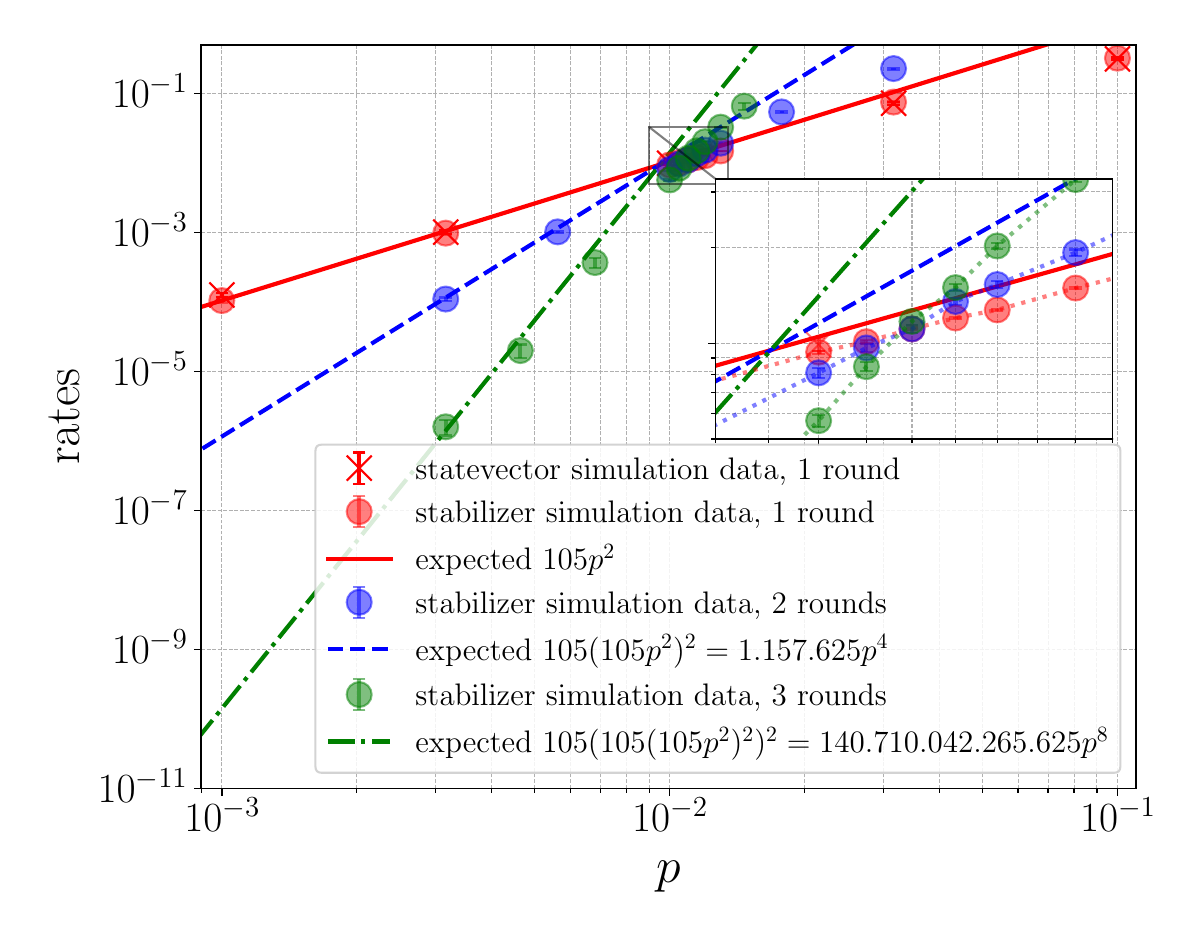}
	\caption{Output error rate of measurement-free 15-to-1 MSD depending on the input magic state error rate $p$. Simulated data points are in good agreement with analytical expressions for $p \lesssim 10^{-2}$. There are $\binom{15}{2} = 105$ uncorrectable \mbox{weight-2} $Z$-errors that determine the leading order of the distillation failure rate of a single round in the low-$p$ limit. Multiple distillation rounds are expected to suppress the failure rate exponentially below the concatenation threshold $p < 1/105$; dotted lines are guides to the eye.}
	\label{fig:performance}
\end{figure}

In the $[[15,1,3]]$ code, there are $\binom{15}{2} = 105$ $Z$-errors that are all uncorrectable and therefore contribute to the logical failure rate $p_L = 105p^2 + \mathcal{O}(p^3)$ in the low-$p$ limit under the noise model in Eq.~\eqref{eq:noise}. We present the results of statevector simulations of the noisy MSD circuit in Fig.~\ref{fig:performance}. The numerical results are in good agreement with the analytical expectation. 

To speed up the numerical simulations, we replace all single-qubit $T$-gates in the physical circuits with identity operations and perform stabilizer simulations. A direct comparison to the statevector simulation results in Fig.~\ref{fig:performance} reveals that we only observe negligible deviations in the resulting output state flip rate. We therefore use stabilizer simulations to numerically estimate the performance of multiple rounds of measurement-free 15-to-1 MSD. It is expected that error rates are exponentially suppressed from $p$ to $c^{-1/t}(c^{1/t}p)^{(t+1)^r}$ for $r$ rounds\footnote{One may as well interpret repeated rounds of MSD with our scheme as a single MSD round based on a concatenated QEC code.} 
of MSD with $c = 105$ and $t = 1$. The numerical simulation results agree with this expectation in a regime where $p \lesssim 4 \times 10^{-3}$ for $r = 3$ rounds of MSD. The crossing point of the analytical error rate curves in Fig.~\ref{fig:performance} is given by $1/c \approx 0.0095$. The simulated data indicates that the threshold lies at approx.~$0.011$, i.e., for $p \lesssim 1\%$ we can expect an improvement of magic state fidelity by sequential magic state distillation. These findings show that three rounds of measurement-free MSD can be sufficient to distill magic states to a noise-level of $\approx 10^{-10}$ with an initial error rate of $p = 10^{-3}$. A fourth round would suppress the noise to about $10^{-18}$.

Finally, we briefly comment on the effect of coherent noise in our MSD scheme. Erroneous coherent rotations about an angle $\theta$ around any Pauli axis $P \in \{X, Y, Z\}$ of the Bloch sphere are described by operators of the form $\exp(-\ii\theta/2P)$. The twirling map relates a coherent rotation channel to a stochastic, incoherent noise map similar to Eq.~\eqref{eq:noise} with an error probability $p = \sin^2(\theta/2) = \mathcal{O}(\theta^2)$ for the error $P$ to occur \cite{katabarwa2015logical, gutierrez2016errors}. Earlier studies hint at a negligible effect of coherent errors in the large-$D$ limit of the underlying logical qubits \cite{beale2018quantum, huang2019performance, iverson2020coherence}. Further investigations could shed light on how measurement-free MSD transforms coherent errors, especially on finite-size logical qubits, and on the interplay with residual noise from finize-size QEC: Since our CFN is capable of correcting any single-qubit Pauli error, arbitrary coherent rotation errors on the input magic states should be permissible in principle. In case one also uses measurement-free QEC routines on the logical qubits \cite{heussen2024measurement, veroni2024optimized}, such that a build-up of coherent errors cannot be prevented by intermediate measurements, we suspect that the twirling procedure would need to be performed \emph{in situ}, e.g.~via randomized compiling \cite{jain2023improved}, in order to achieve the above error rate scaling in practice. 

\textbf{Conclusions and outlook.}
We devise and analyze a novel implementation of 15-to-1 magic state distillation that can be enacted fully deterministically without post-selection and without the need to perform individual qubit measurements. By employing unitary encoding and decoding circuits of the $[[15,1,3]]$ QEC code, we show that a coherent feedback network achieves exponential suppression of noise in magic states upon repeated application of the MSD protocol. Our scheme displays the expected analytical scaling behavior in a regime where the noise-level on magic states is well-below $1\%$, which seems suitable for near-term quantum computing architectures. It may help at simplifying the routing of high-fidelity magic states in real hardware since there occurs no additional overhead associated with (potentially repeated) failure and re-initialization of the distillation procedure.

The design of the coherent feedback network has been performed manually for this work. We anticipate that this process could be automated by classical optimization techniques and might then easily be adapted to other QEC codes \cite{berent2024decoding}. It would be interesting to investigate the scheme on other MSD protocols, especially with more recently discovered code constructions for transversal non-Clifford gates such as \cite{bravyi2012magic, wills2024constantoverheadmagicstatedistillation}. Our concept of unitary decoding with subsequent message qubit corrections could be translated to an \emph{entanglement} distillation scenario \cite{bennett1996purification, pattison2024fastquantuminterconnectsconstantrate, ataides2025constantoverheadfaulttolerantbellpairdistillation}. 

Further studies could jointly investigate measurement-free realizations of finite-distance FT logical qubits combined with concrete magic state distillation sequences and scrutinize the interplay between different sources of noise, for example, to find an advantageous trade-off between the code distance of logical qubits and the required number of distillation rounds to reach a desired overall noise-level.

\section*{Code availability}
All software code used in this project is available from the corresponding author upon reasonable request.

\section*{Acknowledgements}
We appreciate enlightening discussions with Michael Beverland. The encoding circuit in Fig.~\ref{fig:physical_circ} has been synthesized using the Munich Quantum Toolkit, for which we gratefully acknowledge the support of Tom Peham. We thank Nicholas Fazio for useful feedback on the manuscript.

\FloatBarrier
\appendix

\section{Physical circuits for measurement-free 15-to-1 magic state distillation}\label{sec:circ}

\begin{figure*}\centering
	\includegraphics[width=0.99\linewidth]{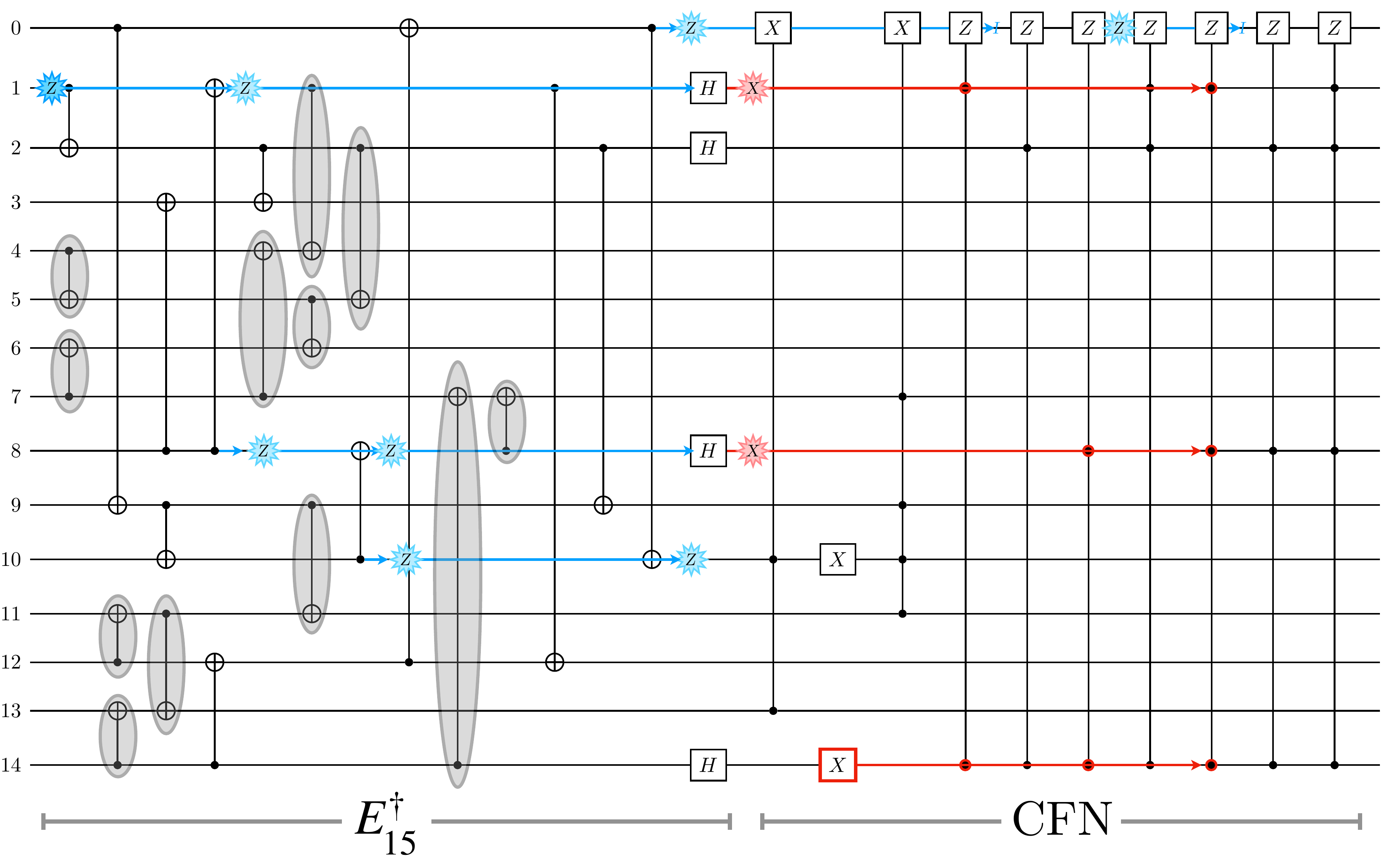}
	\caption{Explicit circuit to perform 15-to-1 MSD (see Fig.~\ref{fig:mdist151}) via the inverse unitary encoding block $E_{15}^\dag$ followed by the coherent feedback network (CFN). The CFN serves to deterministically correct every Pauli error on the output qubit 0, which may result from an arbitrary single-qubit error on any of the 15 input qubits. Qubits 1 to 14 are in the state $\ket{0}$ after $E_{15}^\dag$ in the absence of noise and are discarded at the end of the circuit. As an example, the input error $Z_1$ (blue star) is shown to propagate to qubit 0 while flipping qubits 1 and 8 (see fifth row of Tab.~\ref{tab:zx-sector-flips}) and triggering a combination of three multi-qubit-controlled gates in the CFN (red circles) that corrects the error on the output qubit 0. The multi-qubit-controlled-$X$ gates can be left out when dealing with the noise model of Eq.~\eqref{eq:noise}. The CNOT gates marked gray do not actually propagate logical operations. They can be left out without compromising the ability to distinguish single-qubit $Z$-errors (see Fig.~\ref{fig:physical_circ_cr}).}
	\label{fig:physical_circ}
\end{figure*}

We display the physical circuit for unitary decoding and the coherent feedback network that corrects an arbitrary single-qubit Pauli error in Fig.~\ref{fig:physical_circ}. The multi-qubit-controlled gate sequence is manually crafted to match the conditions for applying Pauli corrections to the output message qubit on wire 0. These conditions are given in Tab.~\ref{tab:zx-sector-flips} and have been found by Pauli propagation of every single-qubit Pauli-$X$ and -$Z$ operator through $E_{15}^\dag$. As an example, propagation of the error $Z_1$ is shown in Fig.~\ref{fig:physical_circ}. After $E_{15}^\dag$, the error has propagated to $Z_0X_1X_8$ indicating that the message qubit suffers a $Z$-flip. Combined with a 0-control on qubit 14, which is compiled with an additional $X$-gate to conventional control-connections that are activated when the input is in the $\ket{1}$ state, this $X$-syndrome triggers an odd number of feedback operations on qubit 0. As a result, the error $Z_0$ is corrected. Note that some input errors do not propagate at all, for instance $Z_{14}$.

\begin{table}[h!]
\centering
\begin{tabular}{c|c||c|c}
\multicolumn{2}{c||}{$Z$-sector} & \multicolumn{2}{c}{$X$-sector} \\ \hline
\textbf{no flip} & \textbf{flip 0} & \textbf{no flip} & \textbf{flip 0} \\ \hline
- & & - & \\
14       & 1             & 3               & 7, 9, 11 \\
1, 14    & 2             & 6               & 10, 11, 12, 13 \\
2, 14       & 8             & 13              & 7, 10, 12, 13 \\
8, 14          & 1, 2          & 5, 6            & \\
1, 2, 14       & 1, 8          & 7, 10           & \\
2, 8, 14          & 2, 8          & 11, 13          & \\
1, 8, 14          & 1, 2, 8       & 3, 5, 9         & \\
1, 2, 8, 14             &              & 4, 5, 6         & \\
              &              & 4, 6, 7         & \\
              &              & 3, 4, 7, 12     & \\
              &              & 7, 9, 10, 11    & \\
              &              & 3, 4, 5, 9, 12  & \\
\end{tabular}
\caption{For the 15 possible single-qubit input errors of $X$- and $Z$-type, only a subset of errors flip the output message qubit 0 while also flipping some of the syndrome qubits 1 to 14. The coherent feedback network in Fig.~\ref{fig:physical_circ} implements these syndrome-conditioned flips while respecting the "no flip" conditions. No feedback can be applied when all syndrome qubits are in the $\ket{0}$ state.}
\label{tab:zx-sector-flips}
\end{table}

\begin{figure*}\centering
	\includegraphics[width=0.8\linewidth]{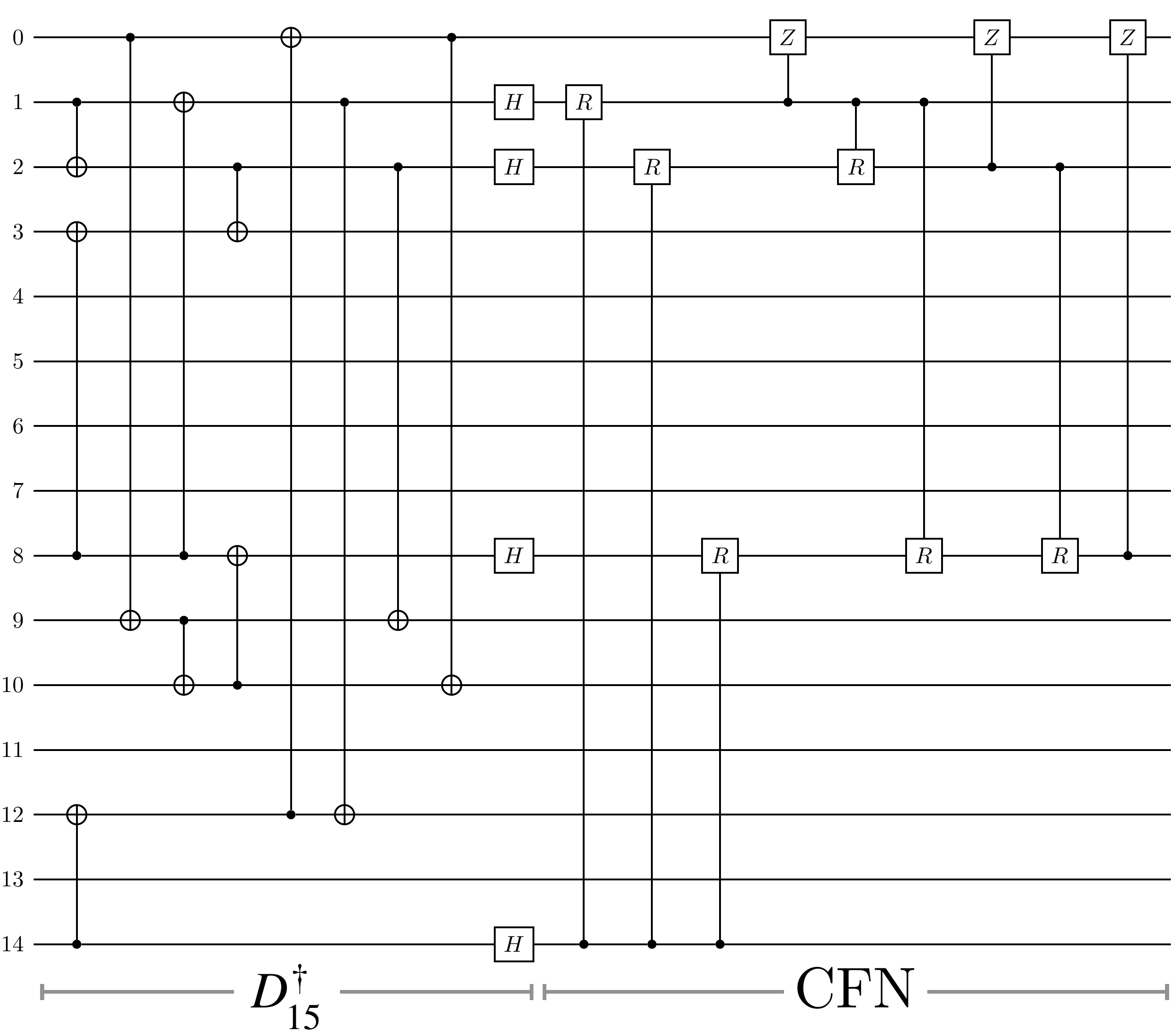}
	\caption{Simplified distillation circuit to correct $Z$-errors only. Unnecessary CNOT gates from $E_{15}^\dag$ in Fig.~\ref{fig:physical_circ} are removed and the resulting circuit is called $D_{15}^\dag$. The CFN can be implemented without multi-qubit-controlled gates if a coherently-controlled-reset ($R$) operation is available. Conditioned on qubit 14, the other three syndrome qubits 1, 2 and 8 are reset. The first controlled-$Z$ operation therefore only corrects qubit 0 in case that syndrome qubits $1, \{1,2\}, \{1,8\}$ or $\{1,2,8\}$ are flipped and then qubits 2 and 8 are reset so that no further feedback is applied. If qubit 1 is in the state $\ket{0}$, the second controlled-$Z$ operation is only applied when the syndrome qubits $2$ or $\{2,8\}$ are flipped and qubit 8 is reset. If also qubit 2 is in the state $\ket{0}$, the last controlled-$Z$ operation only acts if the syndrome qubit $8$ is flipped.
 }
	\label{fig:physical_circ_cr}
\end{figure*}

We note that some CNOT gates in Fig.~\ref{fig:physical_circ}, while mapping stabilizers back to their corresponding physical qubits, need not actually be considered in the context of decoding because they do not propagate physical $Z$-errors to the message qubit 0. Removing these CNOT leaves us with the same correction conditions for the CFN as before: Apply a $Z$-flip to qubit 0 if qubit 1 OR 2 OR 8 are flipped AND qubit 14 is not flipped. If qubit 14 is flipped OR no $X$-syndrome qubits are flipped, apply no $Z$-flip to qubit 0. In Fig.~\ref{fig:physical_circ_cr}, we avoid usage of multi-qubit-controlled gates and instead show an implementation of this CFN by means of a coherently-controlled-reset operation. 

\clearpage

\nocite{*}
\bibliographystyle{mybibstyle}
\bibliography{bibliography}

\end{document}